\documentclass[%
 reprint, 
 amsmath,amssymb,
 aps,
 nofootinbib
 prb,
 superscriptaddress
 ]{revtex4-1}
\usepackage{scrextend}
\usepackage{graphicx}
\usepackage{dcolumn}
\usepackage{bm}
\usepackage{hhline}
\usepackage{tabularx}
\usepackage[utf8]{inputenc}
\usepackage[english]{babel}
\usepackage[bottom]{footmisc}
\usepackage{color}

\begin{document}

\preprint{APS/123-QED}

\title{\textit{Ab initio} study of Bi-based half Heusler alloys as potential thermoelectric prospects.}
\author{Sapna~Singh}\thanks{S.S. and M.Z. contributed equally to this work.} 
\affiliation{Indian Institute of Technology Roorkee, Department of Chemistry, Roorkee 247667, Uttarakhand, India}
\author{Mohd~Zeeshan}\thanks{S.S. and M.Z. contributed equally to this work.}
\affiliation{Indian Institute of Technology Roorkee, Department of Chemistry, Roorkee 247667, Uttarakhand, India}
\author{Jeroen~van~den~Brink}
\affiliation{Institute for Theoretical Solid State Physics, IFW Dresden, Helmholtzstrasse 20, 01069 Dresden, Germany}
\author{Hem~C.~Kandpal}\email{Corresponding author: hem12fcy[at]iitr.ac.in}
\affiliation{Indian Institute of Technology Roorkee, Department of Chemistry, Roorkee 247667, Uttarakhand, India}

\date{\today}
             
\begin{abstract} 
We investigated six heavy element bismuth-based 18-VEC half-Heusler alloys CoTiBi, CoZrBi, CoHfBi, FeVBi, FeNbBi, and FeTaBi by first principles approach, in search of better thermoelectric prospects. The motivation is driven by expected lower thermal conductivity and the recent discovery of CoZrBi-based materials. Significantly, our calculated power factor values of all the systems show an increment of $\sim$40\% in comparison to the reported \textit{p}-type CoTiSb. We propose that doping at Bi-site, on account of electronic features, will be helpful in achieving the proposed power factor values. Interestingly, the thermal conductivity of CoTiBi and CoZrBi was found to be lower and that of CoHfBi was almost parallel, in comparison to the reported CoTiSb. We also provide conservative estimates of the figure of merit, exceeding the reported CoTiSb and comparable to FeNbSb. Overall, our results suggest potential new candidates of bismuth-based ternary compounds for high thermoelectric performance.

\begin{description}
\item[Usage]
Secondary publications and information retrieval purposes.
\item[PACS numbers]
May be entered using the \verb+\pacs{#1}+ command.
\item[Structure]
You may use the \texttt{description} environment to structure your abstract;
use the optional argument of the \verb+\item+ command to give the category of each item. 
\end{description}
\end{abstract}

\pacs{Valid PACS appear here}
\maketitle

\section{Introduction}
Half-Heusler (hH) alloys having 18 valence electron count have been extensively studied over the past few decades in search of better thermoelectric prospects for harvesting the waste heat into electricity \cite{Samanta17, Lee17, Zhao17}. The best performing hH alloys experimentally investigated are stannides and antimonides of cobalt, nickel, and iron, stuffed with the third element from group-IV or group-V of the periodic table, e.g., \textit{M}NiSn, \textit{M}CoSb (\textit{M} = Ti, Zr, Hf) and \textit{M}FeSb (\textit{M} = V, Nb). The \textit{n}-type \textit{M}NiSn, \textit{p}-type \textit{M}CoSb, and \textit{M}FeSb based alloys are reported to have impressive power factor values, however, the performance is restricted by a high intrinsic thermal conductivity \cite{Xie12, Chen13, Berry17}. 

Notwithstanding that great progress has been made in suppressing the thermal conductivity, it could be interesting to investigate the bismuth-based analogues of previously reported hH alloys; from the viewpoint of lower thermal conductivity contribution from heavy elements. While the Bi-doping has been used for improving the transport properties \cite{Barth09, Appel14}, the Bi-based hH alloys have not been explored much. Certainly, it could be challenging to synthesize the ordered compositions of hH alloys with constituents of largely varying atomic mass and melting points. Remarkably, some recent experimental investigations revealed hH alloys comprising heavy atoms such as Ta and Bi, e.g., CoTaSn \cite{Zakutayev13}, CoZrBi \cite{Zhu18}, and FeTaSb \cite{Zhu19}. The discovery of CoZrBi based hH alloys is compelling and could be a benchmark for the synthesis of other Bi-based hH alloys. Most importantly, the pristine CoZrBi was reported to have a much lower thermal conductivity in comparison to the conventional hH alloys \cite{Zhu19}. Therefore, it would be of key importance to investigate the Bi-based hH alloys as potential thermoelectric prospects. In this paper, utilizing \textit{ab initio} approach, semi-classical Boltzmann transport theory, and rigid band approximation, we theoretically investigate the static and dynamical stability, and thermal and electrical transport properties of new Bi-based hH alloys CoTiBi, CoHfBi, FeVBi, FeNbBi, and FeTaBi in cubic $F\bar{4}3m$ symmetry; along with reported CoZrBi. The paper is arranged as follows. 

In Sec. II, we briefly discuss the computational details of structural optimization, electronic structure, transport properties, and approximations used. In Sec. III, we discuss the results of structural optimization, static and dynamic stability (phonons), electronic structure (band structure and DOS), and electrical and thermal transport properties. A summary of the paper is given in Sec. IV. Throughout this paper, we use the term Co-group for CoTiBi, CoZrBi, CoHfBi, and Fe-group for FeVBi, FeNbBi, FeTaBi systems.

\section{Computational Details}
We used a combination of two different first-principles density functional theory (DFT) codes: the full-potential linear
augmented plane wave method (FLAPW) \cite{Singh06} implemented in Wien2k \cite{Blaha01} and the plane-wave pseudopotential
approach implemented in Quantum Espresso package \cite{Giannozzi09}. The former has been used to obtain equilibrium lattice
constants, electronic structure, and transport properties, and the latter to confirm the structure stability by determining
the phonon spectrum. 

The FLAPW calculations were performed using a modified Perdew-Burke-Ernzerhof (PBEsol correlation) \cite{Perdew08} implementation
of the generalized gradient approximation (GGA). The Becke-Johnson potential \cite{Becke06} was used for band structure calculations.
For all the calculations, the scalar relativistic approximation was used. We also performed full relativistic calculations
to see the effect of spin-orbit coupling. Nonetheless, the spin-orbit coupling has not much effect on the electronic structures close to
the Fermi level. Therefore, we performed electronic structure calculations under scalar relativistic approximation for all the systems.
The muffin-tin radii (RMTs) were taken in the range 2.32--2.50 Bohr radii for all the atoms. RMT {$\times$} kmax = 9 was set as the
plane wave cutoff. The self-consistent calculations were employed using 125000 \textit{k}-points in the full Brillouin zone.
The energy and charge convergence criterion was set to 10$^{-6}$ Ry and 10$^{-5}$ e, respectively. 

The electrical transport properties have been calculated using the Boltzmann theory \cite{Allen} and relaxation time approximation
as implemented in the Boltztrap code \cite{Madsen06}. The Boltztrap code utilizes input from Wien2k code. The electrical conductivity
and power factor are calculated with respect to time relaxation, \textit{$\tau$}; the Seebeck coefficient is independent of
\textit{$\tau$}. The relaxation time was calculated by fitting the available experimental data with theoretical data. We have
used this approach in evaluating the electronic transport properties of our systems.  

In the plane-wave pseudopotential approach, we used scalar-relativistic, norm-conserving pseudopotentials for a plane-wave
cutoff energy of 100~Ry. The exchange-correlation energy functional was evaluated within the GGA, using the Perdew-Burke-Ernzerhof
parametrization \cite{Perdew96}, and the Brillouin zone was sampled with a 20$\times$20$\times$20 mesh of Monkhorst-Pack \textit{k}-points.
The calculations were performed on a 2$\times$2$\times$2 \textit{q}-mesh in the phonon Brillouin zone.  

The lattice thermal conductivity was obtained by solving linearized Boltzmann transport equation (BTE) within the single-mode
relaxation time approximation (SMA) \cite{Ziman1960} using thermal2 \cite{thermal2} code as implemented in the Quantum Espresso
package. We employed generalized gradient approximation (GGA) given by Perdew-Burke-Ernzerhof (PBE) \cite{Perdew96} for
exchange-correlation functional. We have chosen Troullier-Martins norm-conserving pseudopotentials from the Quantum Espresso
webpage \cite{QEweb}. An energy cutoff of 100~eV was used for the plane-wave basis set and Brillouin zone integration was
performed on a Monkhorst-Pack 20$\times$20$\times$20.

\section{Results}

\subsection{Structural Optimization and Phonon Stability}
The ground state properties of all the systems were calculated by GGA-PBEsol implemented in Wien2k. The approximation is reliable for calculating the ground state properties \cite{He14}. The structure of hH alloys crystallize in MgAgAs-type structure and can be seen as a stuffed combination of rock salt and zinc blende structure \cite{Graf11}. The structure of all six \textit{XYZ} hH alloys
was optimized in cubic \textit{$F\bar{4}3m$} symmetry, where \textit{X} = Co, Fe, \textit{Y} = Ti, Zr, Hf, V, Nb, Ta, and \textit{Z} = Bi. For optimization, fitted with Birch-Murnaghan equation of state \cite{Birch47}, the total energy was minimized as a function of volume for each system. The optimized lattice parameters and band gap values are listed in Table~\ref{optimization}. 

\begin{table}[]
\caption{The optimized lattice constant \textit{a} and band gap \textit{E$_g$} values of \textit{XYZ} half-Heusler alloys (\textit{X} = Co, Fe \textit {Y} = Ti, Zr, Hf, V, Nb, Ta, and \textit{Z} = Bi) in cubic \textit{$F\bar{4}3m$} symmetry.}
\centering
\setlength{\arrayrulewidth}{0.5pt}
\begin{tabular*}{\columnwidth}{c @{\extracolsep{\fill}} lcc}
\hline 
\hline
System      & \textit{a} (\AA)  & \textit{E$_g$} (eV) \\ \hline
CoTiBi      & 5.9497          	& 0.84 \\
CoZrBi      & 6.1357          	& 1.00 \\
CoHfBi      & 6.1076          	& 0.91 \\ \hline
FeVBi       & 5.8603          	& 0.48 \\
FeNbBi      & 6.0130          	& 0.64 \\
FeTaBi      & 6.0077          	& 0.97 \\ 
\hline
\hline
\end{tabular*}
\label{optimization}
\end{table}

The calculated lattice parameter of CoZrBi is in fairly good agreement with the experimental value of 6.18 \AA \cite{Zhu18}. The trend of lattice parameter is consistent for both Co-group and Fe-group, first increasing from first to the second member, followed by a slight decrease in the third member on account of lanthanide contraction. Importantly, the band gap survives in all the cases and the values range 0.4--1.0 eV, the lowest and highest values are for FeVBi and CoZrBi, respectively. The calculated band gap in case of CoZrBi is in agreement with previously calculated values \cite{Fiedler16}. No experimental band gap value of CoZrBi is reported for comparison. Band gap values are consistently increasing for Fe-group, however, the band gap value of CoHfBi is slightly lower than the CoZrBi. Nevertheless, the existence of band gap in all the cases is more important as semiconductors are the best choices for thermoelectric materials. Following this, the optimized ground state structures were studied for dynamic stability with phonon calculations.

\begin{figure}
\centering\includegraphics[scale=0.4]{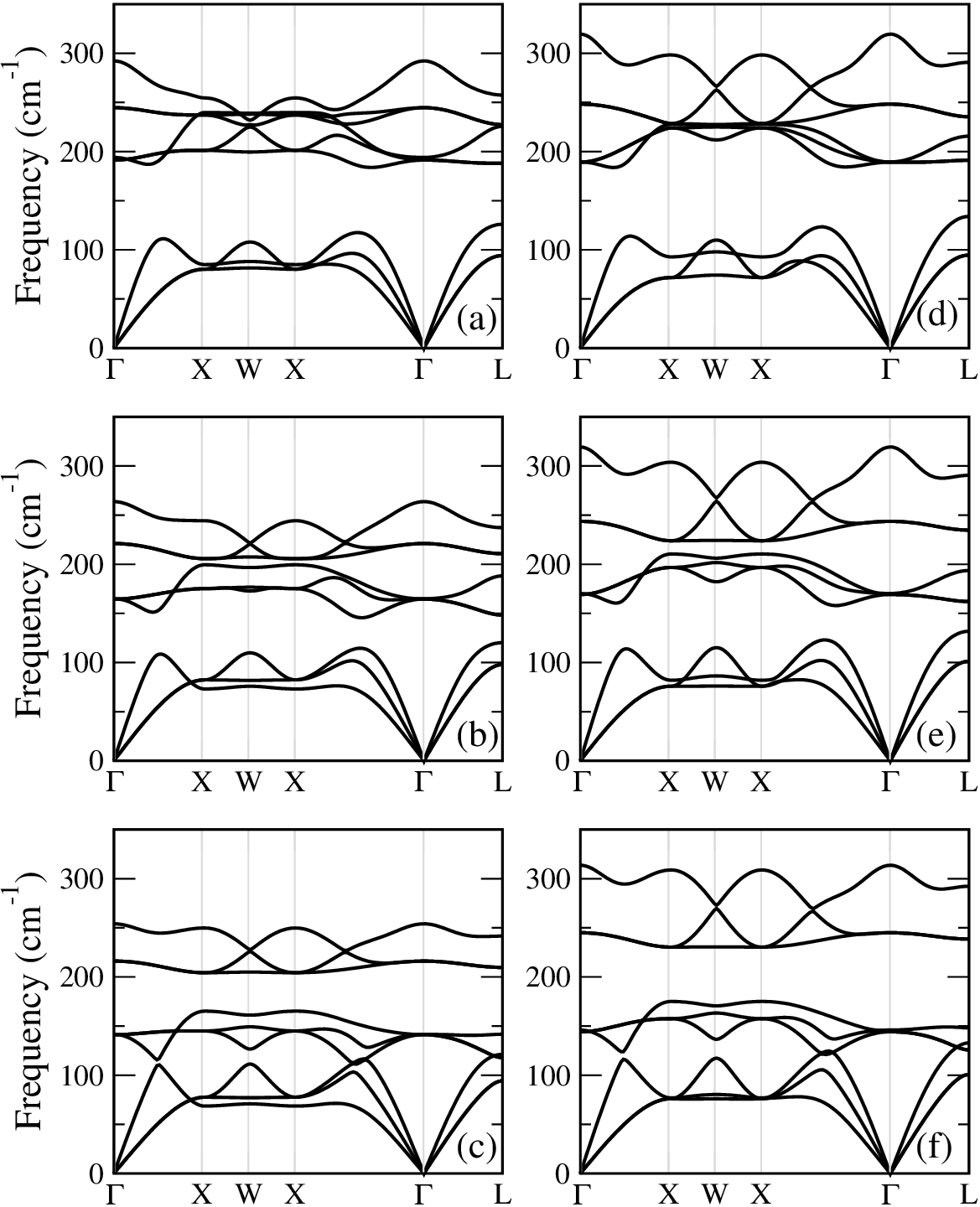}
\caption{Phonon dispersion plots of a) CoTiBi, b) CoZrBi, c) CoHfBi, d) FeVBi, e) FeNbBi, and f) FeTaBi in cubic \textit{$F\bar{4}3m$} symmetry.}
\label{phonon}
\end{figure}

Phonon calculations were performed by Quantum Espresso -- based on DFT and plane-wave pseudopotential method. It is a two-step procedure. The first step is the optimization of ground state structure, followed by the calculation of phonon dispersions utilizing density functional perturbation theory (DFPT) implemented in Quantum Espresso. The calculations were performed on 2$\times$2$\times$2 mesh in the phonon Brillouin zone and force constants in real space derived from this input are used to interpolate between \textit{q} points and to obtain the continuous branches of the phonon band structure. For dynamical stability, the phonon frequencies should be real and not imaginary \cite{Elliott06, Togo15}. We observe from Fig.~\ref{phonon} that all the proposed systems have the real phonon frequencies throughout the Brillouin zone, validating their dynamic stability in cubic \textit{$F\bar{4}3m$} symmetry. The survival of band gap and dynamical stability of proposed systems prompt us to investigate their electronic structure and transport properties. 

\subsection{Electronic Structure}
\begin{figure}
\centering\includegraphics[scale=0.5]{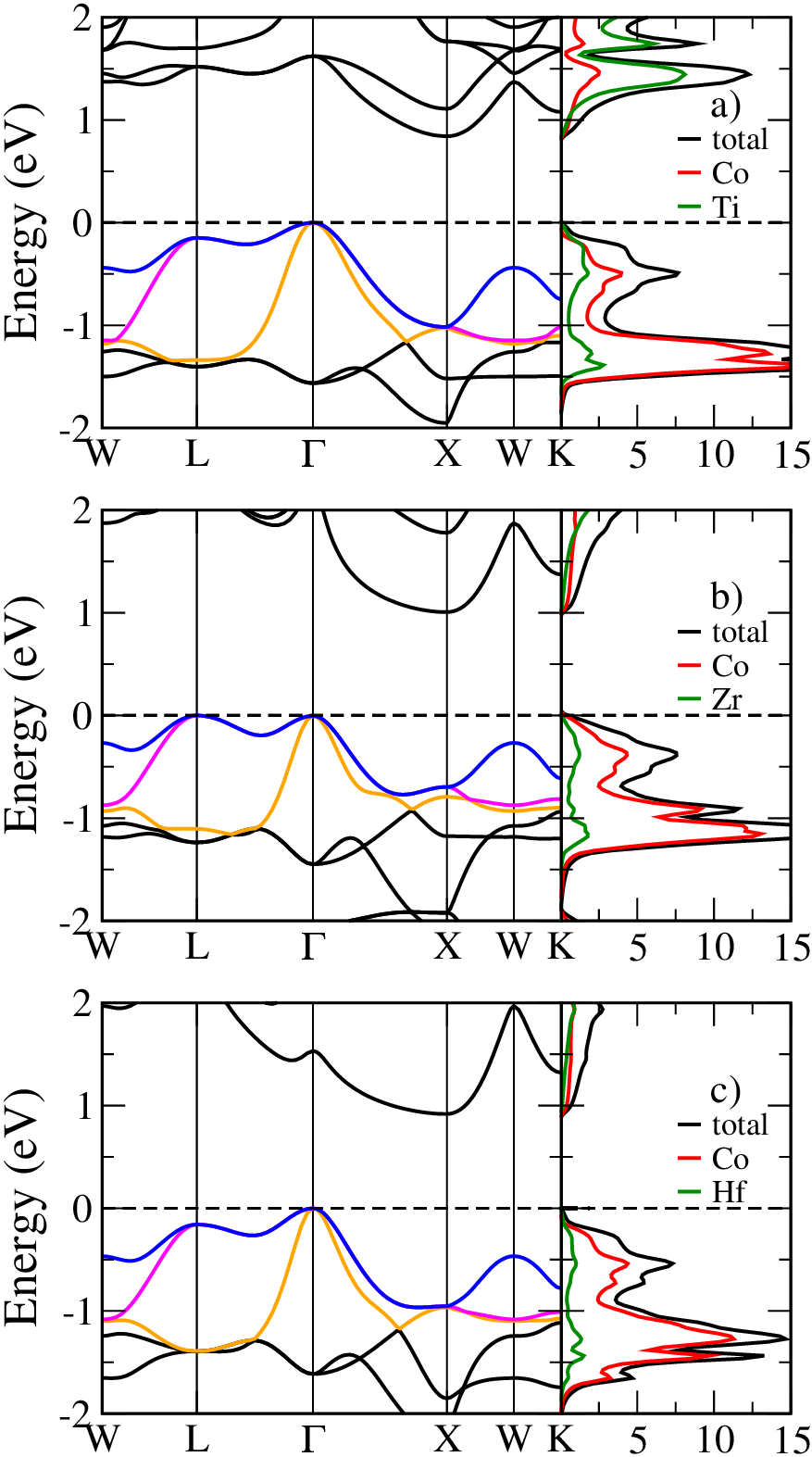}
\caption{Electronic bands and density of states of a) CoTiBi, b) CoZrBi, and c) CoHfBi in cubic \textit{$F\bar{4}3m$} symmetry. The top of the valence band is taken as zero on the energy axis.}
\label{band_I}
\end{figure}

\begin{figure}
\centering\includegraphics[scale=0.5]{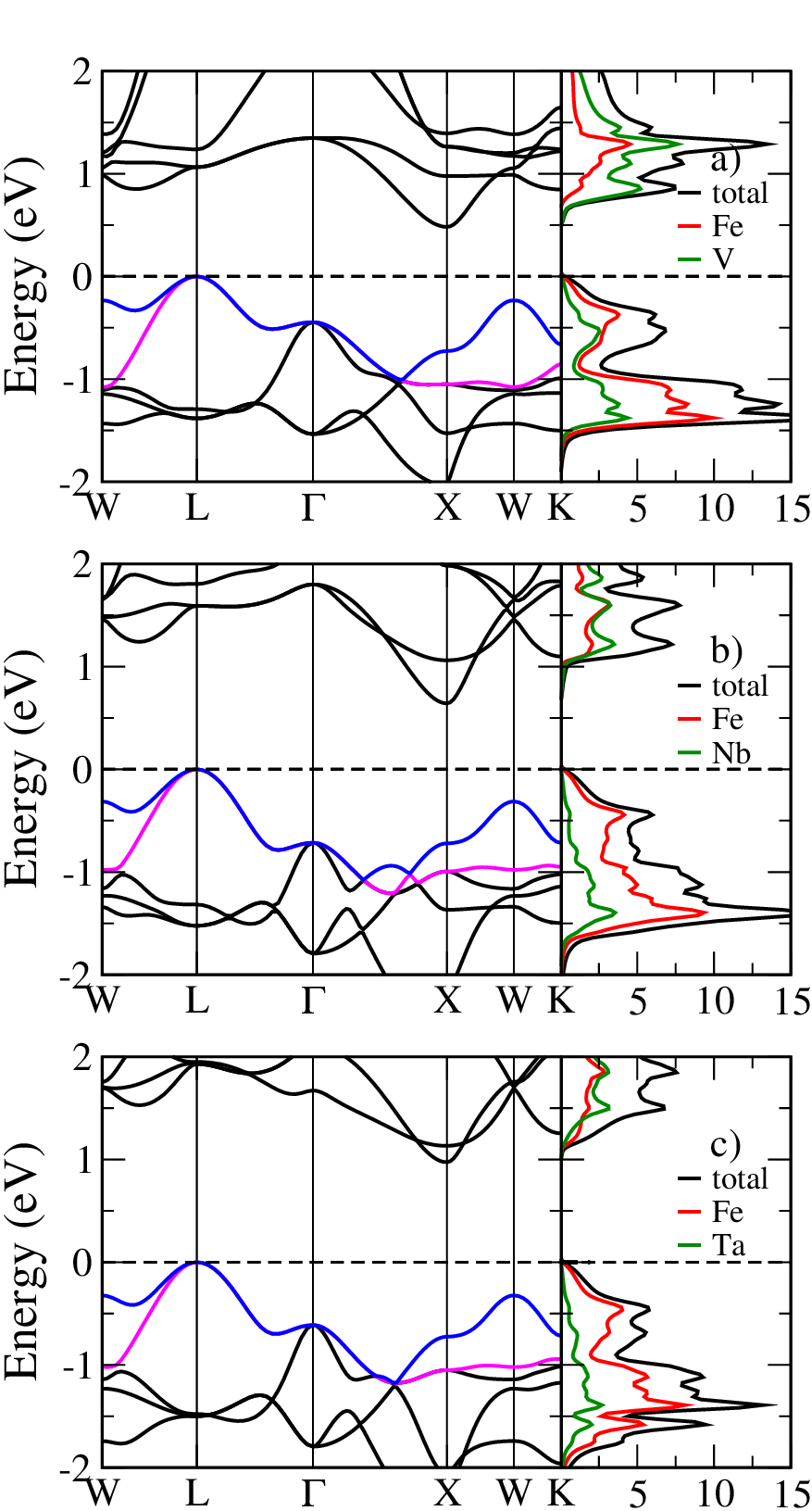}
\caption{Electronic bands and density of states of a) CoTiBi, b) CoZrBi, and c) CoHfBi in cubic \textit{$F\bar{4}3m$} symmetry. The top of the valence band is taken as zero on the energy axis.}
\label{band_II}
\end{figure}

The electronic structures of all the proposed systems were calculated by Becke-Johnson potential \cite{Becke06} implemented in Wien2k. Figure~\ref{band_I} and~\ref{band_II} shows the electronic band structure and density of states (DOS) of Co-group and Fe-group, respectively. The valence band maximum (VBM) of Co-group is located at $\Gamma$ point for CoTiBi and CoHfBi, and at L point for CoZrBi. The conduction band minimum (CBM) of these systems is located at X point. In case of Fe-group, the VBM and CBM of all three systems are located at L point and X point, respectively. Thus, all the systems are indirect band gap semiconductors. The VBM of Co-group and Fe-group is threefold and twofold degenerate, respectively. The degeneracy plays an important role in governing the transport properties. The combination of heavy and lights bands at VBM facilitates the electrical transport properties on \textit{p}-type doping. Where heavy bands improve the Seebeck coefficient on account of heavy effective mass, the light bands facilitates the mobility of charge carriers \cite{Gudelli13, Zhang10, Yang14, Li16}. 

In case of Co-group, the electronic structures of CoTiBi and CoHfBi are quite similar in the valence band region and likely to exhibit similar transport properties. The bands at VBM are relatively flat in CoZrBi as compared to CoTiBi and CoHfBi, indicative
of a higher Seebeck coefficient in CoZrBi. Interestingly, in the case of CoZrBi, the energy difference between the VBM band (L point) and VBM-1 band ($\Gamma$ point) is only 0.004 eV. Thus, both L and $\Gamma$ points can almost equally contribute to the charge carrier transport. The higher band degeneracy near the VBM suggests that CoZrBi could be a more promising thermoelectric material as compared to CoTiBi and CoHfBi. Further, the higher band degeneracy near the VBM in CoZrBi could be responsible for a higher power factor in comparison to well-known CoTiSb, as noted by Zhu \textit{et al} \cite{Zhu18}. The electronic band structures of Fe-group are quite similar. The resembling electronic features of FeNbBi and FeTaBi can be attributed to the similar sizes and chemical properties of Nb and Ta.

The DOS features show that the VBM of Co-group is mostly populated by the \textit{d}-states of Co and some states from \textit{Y}
(= Ti, Zr, Hf). The CBM is almost equally populated by the \textit{d}-states of Co and \textit{Y} atom in case of CoZrBi and CoHfBi whereas the Ti-states mostly dominate in case of CoTiBi. Similar to Co-group, the VBM of Fe-group is mostly occupied by \textit{d}-states of Fe and some states from \textit{Y} (= V, Nb, Ta) and the CBM is almost equally occupied by \textit{d}-states of Fe and \textit{Y} atom. Notably, no significant contribution was observed from Bi either in VBM or CBM in all the systems, not shown for clarity. This suggests that any doping at Bi-site for suppressing the lattice thermal conductivity will be most effective as it would not affect the underlying band structure much, i.e., intact electrical transport properties. 

This assertion may be validated by comparing the two different experimental works on CoZrBi. Recently, Zhu \textit{et al.} doped CoZrBi at Bi-site by Sn and obtained impressive power factor values. The room temperature power factor value was reported to be 25 $\mu$W cm$^{-1}$ K$^{-2}$ and the peak value obtained was 40 $\mu$W cm$^{-1}$ K$^{-2}$ \cite{Zhu18}. On the other hand, Ponnambalam \textit{et al.} \cite{Ponnambalam07} reported low power factor values when doped CoZrBi at Co-site by Ni. As discussed, the doping
at Co-site by Ni might have altered the high band degeneracy near the VBM as Co-states contribute mostly to it. Another possibility could be the \textit{n}-type doping contrary to the favorable \textit{p}-type doping, as predicted by electronic band structure. 

Thus, we believe that the electrical transport properties of the proposed Bi-based hH alloys will be much improved on \textit{p}-type doping at Bi-site. The electronic bands favorable for good transport properties in all the systems correspond to the experimentally realizable doping levels, as discussed in the next section.

\subsection{Transport Properties}
In this section, we calculate the transport properties of six Bi-based hH alloys in cubic \textit{$F\bar{4}3m$} symmetry.
Thermoelectric efficiency of a material is determined by a dimensionless quantity called the figure of merit, given by $ZT=S^2\sigma T/\kappa (\kappa = \kappa_e + \kappa_l)$, where S is the Seebeck coefficient, $\sigma$ is the electrical conductivity, and $\kappa$ is the thermal conductivity comprising the electronic and lattice parts \cite{Sevincli13, Boona17, Tan17}. Electrical transport coefficients \textit{S}, $\sigma$, and thereby $S^2 \sigma$, are calculated by Boltztrap code within the rigid band approximation (RBA). Further, the constant relaxation time approach (CRTA) is used for evaluating the transport coefficients. Within this approach, the Seebeck coefficient is taken to be independent of relaxation time, $\tau$, and electrical conductivity and power factor (PF), $S^2 \sigma$, are expressed with respect to the $\tau$. The RBA and CRTA have been used successfully by many groups for theoretical designing of new thermoelectric materials \cite{Jodin04, Chaput05, Madsen06JACS, Yang08, Lee11}. The thermal transport coefficient $\kappa_e$ is calculated by Boltztrap code whereas the $\kappa_l$ is calculated by thermal2 code implemented in Quantum Espresso. 

\begin{figure}
\centering\includegraphics[scale=0.4]{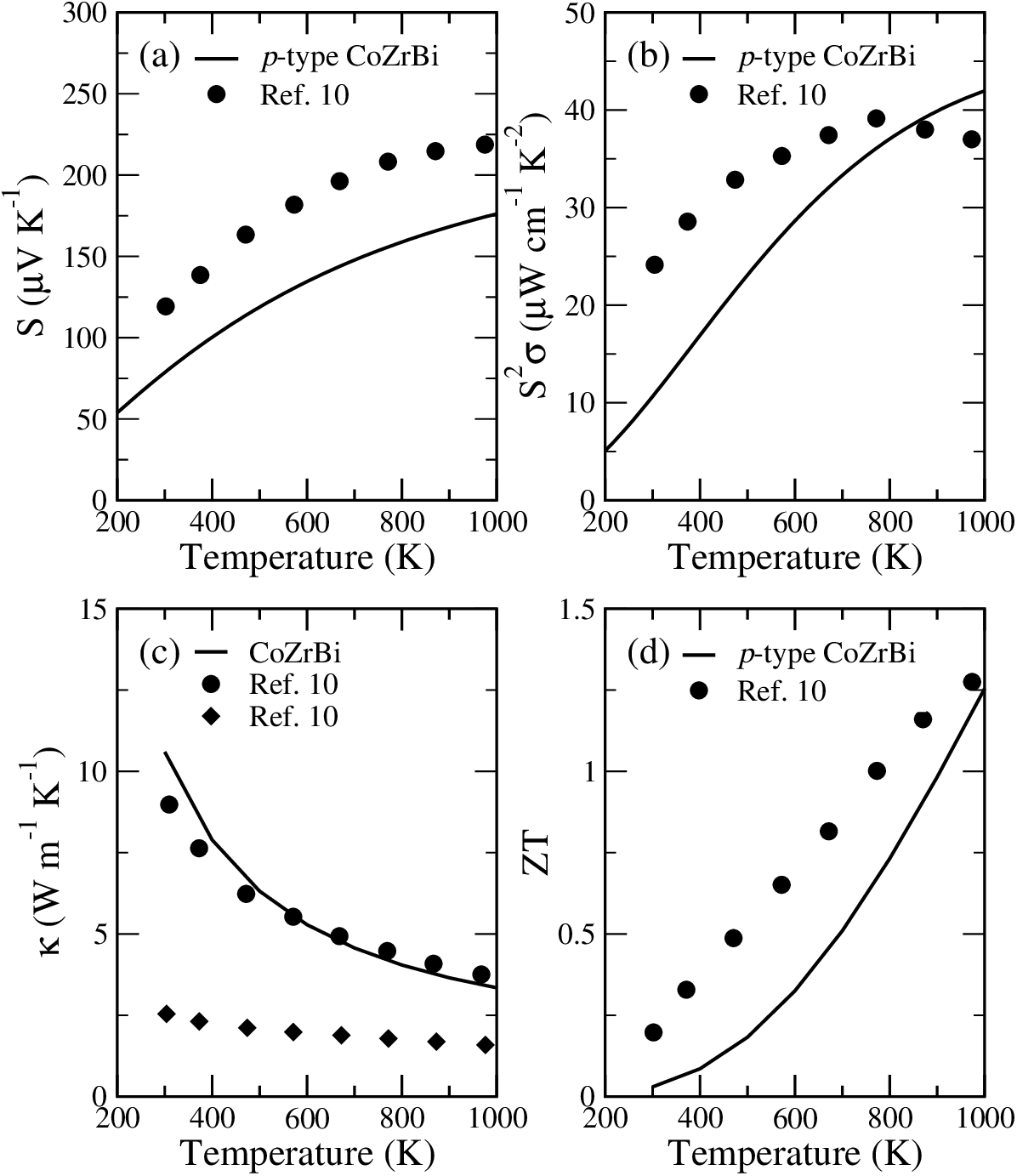}
\caption{a) Seebeck coefficient, b) Power factor, and d) ZT as a function of temperature for calculated 0.24 \textit{p}-type doped CoZrBi and reported 0.20 \textit{p}-type doped CoZrBi. c) Thermal conductivity as function of temperature for calculated parent CoZrBi, and reported parent CoZrBi and 0.20 \textit{p}-type doped CoZrBi.}
\label{expt}
\end{figure}

We, before discussing our results on the proposed systems, corroborate our calculated values with the experimentally reported CoZrBi. The increasing trend of calculated and reported Seebeck coefficient for \textit{p}-type CoZrBi is illustrated in Fig.~\ref{expt}a. The calculated \textit{S} is of 0.24 \textit{p}-type doped CoZrBi whereas the reported \textit{S} is of 0.20 \textit{p}-type doped CoZrBi, i.e., CoZrBi$_{0.80}$Sn$_{0.20}$ \cite{Zhu18}. Hole doping of 0.24 per unit cell is the optimal doping level obtained for attaining maximum PF. The calculated and reported optimal doping levels are in fairly good agreement. The slight variation in values is accepted and likely to vary with the experimental doping concentrations, as discussed ahead. Our calculated values of \textit{S} are slightly underestimated as compared to the experimental ones. This can be attributed to the varying calculated and experimental doping levels. Nevertheless, the trend of the two plots is more important and encouraging for us. 

Curiously, we compare our calculated PF values with the experimentally reported values of CoZrBi. As was mentioned, within the CRTA, the PF is generally expressed with respect to constant relaxation time. In order to have a more realistic evaluation of PF values, instead of presenting $S^2\sigma/\tau$, we present $S^2\sigma$ values utilizing a constant relaxation time, $\tau$ = 2$\times10^{-15}$~s, for all the proposed systems. The chosen constant number is not random, indeed, obtained by careful analysis of calculated and experimental values of electrical conductivity of CoZrBi. The strategy has been helpful in our previous works. In a crude approximation, employing $\tau$ = $\sigma_{exp}/\sigma_{cal}$,
we found that $\tau$ was of the order of 10$^{-15}$~s for CoTiSb and FeNbSb. Utilizing $\tau$ = 10$^{-15}$~s for CoTiSb and
$\tau$ = 2$\times10^{-15}$~s for FeNbSb, we obtained a nice agreement between calculated and reported PF values. The experimentally reported PF of \textit{p}-type CoTiSb was $\sim$23 $\mu$W cm$^{-1}$ K$^{-2}$ at 850~K whereas our calculated  value was $\sim$20 $\mu$W cm$^{-1}$ K$^{-2}$ at 900~K \cite{Zeeshan17}. Similarly, our calculated PF of $\sim$40 $\mu$W
cm$^{-1}$ K$^{-2}$ of \textit{p}-type FeNbSb at 1100~K was close to the reported PF of $\sim$45 $\mu$W cm$^{-1}$ K$^{-2}$ at 1100~K \cite{Zeeshan18}. 

Optimistic of our approach, in a similar manner, we found that $\tau$ is of the order of 10$^{-15}$~s for \textit{p}-type CoZrBi and we take a constant value of relaxation time, i.e., $\tau$ = 2$\times10^{-15}$~s, to compare our calculated values with the experiment. The increasing trend of PF as a function of temperature for calculated and reported \textit{p}-type CoZrBi is depicted in Fig.~\ref{expt}b. Initially, the experimental PF values dominate the calculated values, anyhow, the two plots approach each other at higher temperature. Most importantly, the trend of the two plots is same. The close proximity at high temperature is further motivating as we are interested in high temperature thermoelectric applications. Having obtained a reasonable agreement of thermopower and power factor, next, we bring into comparison the calculated and reported thermal conductivity of CoZrBi.

Figure~\ref{expt}c indicates the decreasing trend of thermal conductivity as a function of temperature of calculated and reported parent CoZrBi, and reported \textit{p}-type CoZrBi. Total $\kappa$ comprises two components $\kappa_e$ and $\kappa_l$, the former is obtained by Boltztrap code where the latter was obtained from Quantum Espresso code. The total $\kappa$ is dominated by $\kappa_l$ and is well-known for hH alloys \cite{Kimura08, Yu17}. It has been our observation in previous two works that the $\kappa_l$ values computed by thermal2 code for ternary hH alloys are tenfold higher and we obtained a nice agreement for CoTiSb \cite{Zeeshan17_II} and FeNbSb \cite{Zeeshan18}. Making use of this observation, we obtain a nice agreement between the calculated and reported $\kappa$ of parent CoZrBi, further substantiating our assumption. Unfortunately, as of now, it was not possible for us to calculate the $\kappa$ of doped systems considering the expensive computational requirements.

Therefore, we resort to utilizing $\kappa$ of parent undoped systems for the evaluation of \textit{ZT} values (= $S^2\sigma T/\kappa$). The numerator part, $S^2\sigma$, is calculated for doped systems, albeit the $\kappa$ of parent system is incorporated. Here, we stress that our calculated \textit{ZT} values are likely to be underestimated as $\kappa$ reduces with doping. This can be seen from the reported $\kappa$ of parent and doped CoZrBi, Fig.~\ref{expt}c. At room temperature, the difference in the $\kappa$ values of parent and doped CoZrBi is significant but the two plots approach each other in the high temperature region. This suggests that our calculated \textit{ZT} values may not be far from the actual values and represent a more realistic guess to the experimentalists. 

The figure of merit \textit{ZT} as a function of temperature is shown in Fig.~\ref{expt}d. Inevitably, resembling the trend of PF, the \textit{ZT} values of reported \textit{p}-type CoZrBi dominates our calculated values. Yet again, the two plots approach each other at high temperatures. Overall, we obtained a satisfactory agreement of thermopower, power factor, thermal conductivity, and \textit{ZT} values between the calculated and reported values of CoZrBi. The magnitude may vary more or less, however, the trend of the two plots is more important to us. The deviations in magnitude are well expected as our calculated values are for the pristine system under ideal conditions, very unlikely to be the case of experimental study. Further, our calculated values are for \textit{p}-type CoZrBi, in general, rather than considering a particular atom for doping. The transport properties may vary depending on the choice of dopant. Wu \textit{et al.} obtained a PF of 9 $\mu$W cm$^{-1}$ K$^{-2}$ at 850~K on 15\% \textit{p}-type doping of Sb by Ge \cite{Wu09}. However, the same group obtained a high PF of 23 $\mu$W cm$^{-1}$ K$^{-2}$ at 850~K on 15\% \textit{p}-type doping of Co by Fe \cite{Wu07}. This shows that at the same working temperature and same doping levels, the PF values may vary significantly depending on the choice of dopant. 

The above discussion and the agreement between our calculated and reported values show the reliability of our calculations and how realistic are the proposed values. Moreover, our calculated values are very close to the experimental values in the high temperature range. This is favorable for us as we are proposing our results for the high temperature thermoelectric applications. Motivated by the corroboration between calculated and experimental values for CoZrBi, we proceed next to study the transport properties of the proposed systems. We justified the use of $\tau$ = 2$\times$10$^{-15}$ s as an appropriate constant relaxation time for CoZrBi, hence, can be used for the Co-group.  We use the same value for Fe-group also. This is because no experimental study is reported on Fe-based bismuth containing hH alloys and we aid our previous work on FeNbSb, where we used the same value of relaxation time \cite{Zeeshan18}. 

\begin{figure}
\centering\includegraphics[scale=0.45]{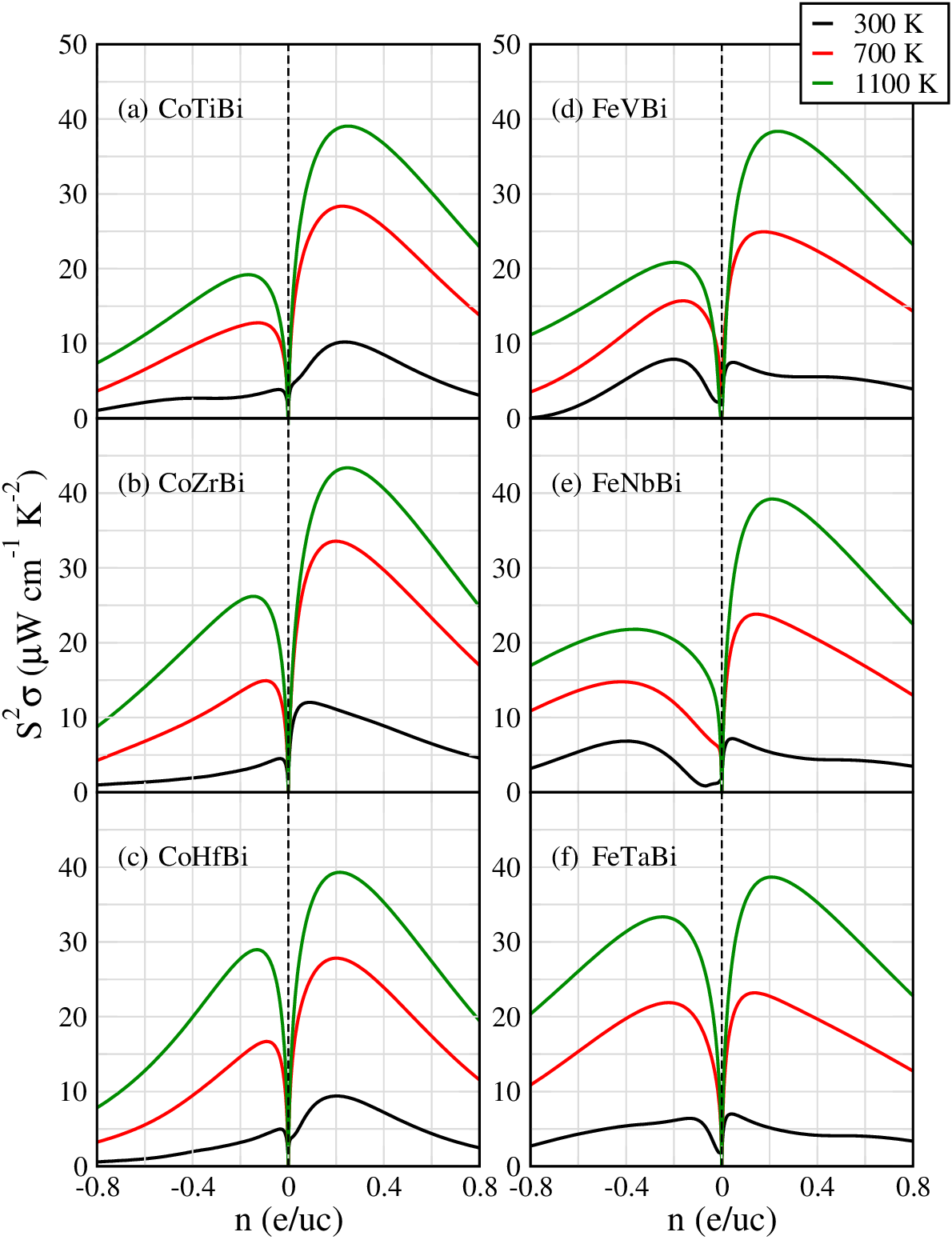}
\caption{Power factor as a function of doping of \textit{XYZ} half-Heusler alloys (\textit{X} = Co, Fe, \textit {Y} = Ti, Zr, Hf, V, Nb, Ta, and \textit{Z} = Bi) at 300, 700, and 1100 K in cubic \textit{F$\bar4$3m} symmetry, assuming relaxation time $\tau$ = 2$\times$10$^{-15}$ s. The \lq{+\rq} and \lq{--\rq} signs on the horizontal axes represent the hole and electron doping, respectively.}
\label{PF}
\end{figure}

The PF as a function of doping at 300, 700, and 1100~K for Co-group and Fe-group, assuming $\tau$ = 2$\times$10$^{-15}$ s,
is shown in Fig.~\ref{PF}. The trend of PF is consistent in all the systems, initially, the PF increases with either \textit{n}-type or \textit{p}-type doping and then decreases gradually at high doping levels. However, \textit{p}-type doping dominates in all the cases. The trend can be understood in terms of the effect of doping on Seebeck coefficient and electrical conductivity. Seebeck coefficient is large and electrical conductivity is low when the Fermi level is near the middle of the band gap. On either type of doping, the Fermi level moves toward valence band or conduction band leading to decrease of Seebeck coefficient, on the other hand, the electrical conductivity increases as the carrier concentration increases with doping. 

The two conflicting trends of Seebeck coefficient and electrical conductivity demands an optimal doping concentration at which the maximum PF can be obtained. Generally, such a doping level is found close to the band edge. This is what we have observed in all cases. The prediction of optimal doping levels enables the experimentalist to target a specific range of doping levels while exploring new materials. 

The PF values at 300~K are the lowest and the values improve with temperature, i.e., the highest values are obtained at 1100~K. Our motive is to study the proposed hH alloys for high temperature thermoelectric applications, here onwards, we focus on the values at 1100~K. The transport properties of Co- and Fe-based hH alloys are experimentally reported at high temperatures such as 900--1100~K which suggests that these alloys may sustain such high temperatures \cite{Sekimoto05, Sekimoto06, Fu15}. At 700~K, an interesting observation for FeTaBi is the comparable values of PF at both \textit{p}-type and \textit{n}-type doping. Ideally, for a good thermoelectric performance, both \textit{p}-type and \textit{n}-type legs of a thermoelectric module should be of same or similar materials, which is rare \cite{Dubois_Book}. This suggests the potential of FeTaBi at 700~K as both \textit{p}-type and \textit{n}-type thermoelectric material. 

At 1100~K, in Co-group, the PF of CoZrBi is higher than both CoTiBi and CoZrBi on account of high band degeneracy. Surprisingly, the values of CoTiBi and CoHfBi are not far behind. This can be attributed to their high electrical conductivity as compared to CoZrBi, Table~\ref{zT}. For Fe-group, as anticipated, the PF values of FeVBi, FeNbBi, and FeTaBi are quite similar on account of almost identical electronic features, as discussed in the previous section. The PF values of CoTiBi, CoHfBi, and Fe-group are almost competitive to CoZrBi which suggests the thermoelectric potential of the proposed systems. It will be interesting to see whether the thermal conductivity of the proposed systems is as low as that of CoZrBi.

\begin{figure}
\centering\includegraphics[scale=0.45]{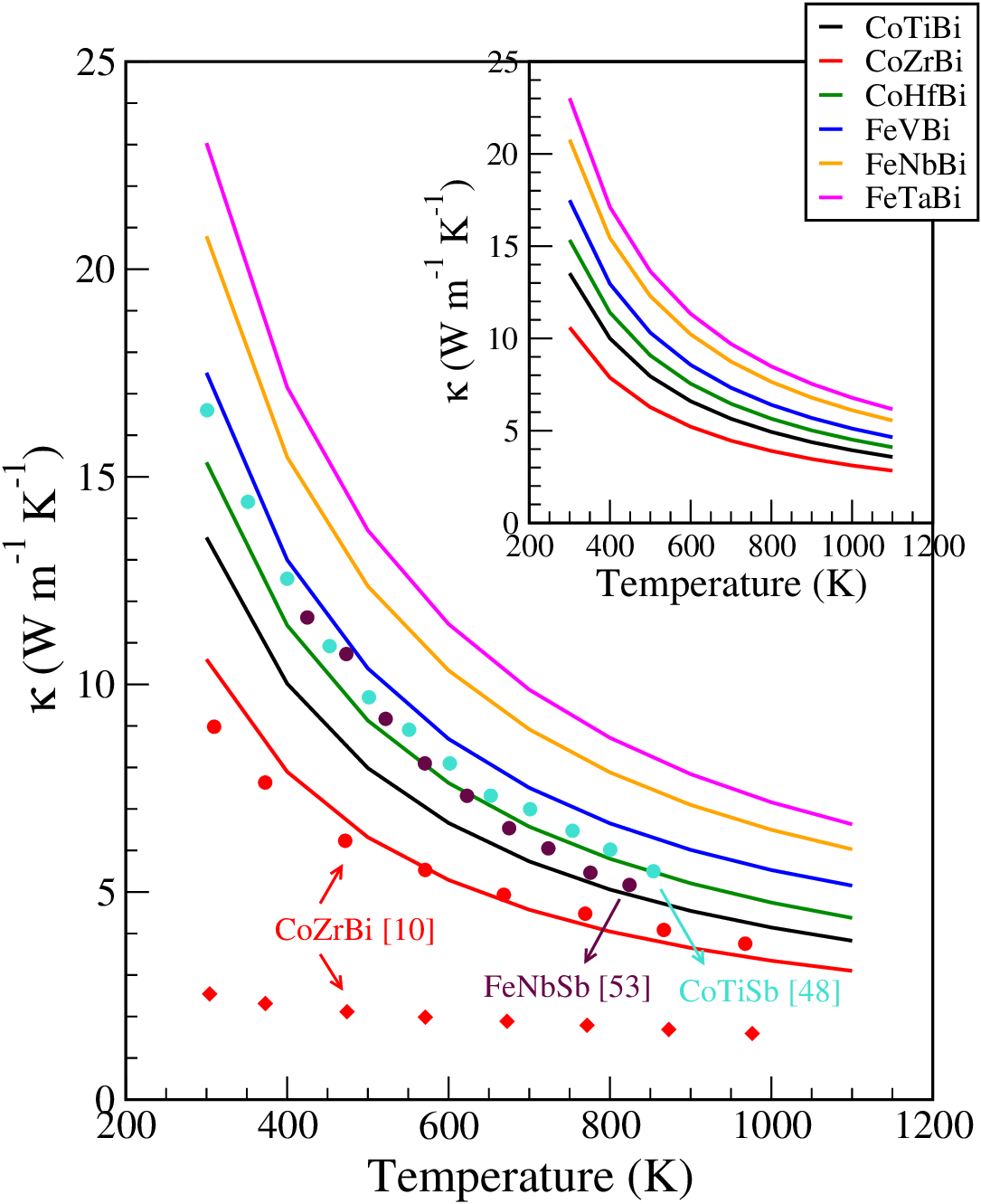}
\caption{Calculated thermal conductivity as a function of temperature of \textit{XYZ} half-Heusler alloys (\textit{X} = Co, Fe, \textit {Y} = Ti, Zr, Hf, V, Nb, Ta, and \textit{Z} = Bi) in cubic \textit{F$\bar4$3m} symmetry. The reported thermal conductivity of parent CoTiSb, FeNbSb, CoZrBi, and 0.20 \textit{p}-type CoZrBi are also shown. Inset shows the calculated lattice thermal conductivity as a function temperature of the proposed systems. The circles and diamonds represent parent and doped compositions, respectively.}
\label{kappa}
\end{figure}

Figure~\ref{kappa} shows the decreasing trend of thermal conductivity as a function of temperature for Co-group and Fe-group, along with the reported thermal conductivity of parent CoTiSb, FeNbSb, CoZrBi, and \textit{p}-type CoZrBi. The inset shows the dominance of lattice thermal conductivity in total thermal conductivity. The calculated $\kappa$ of CoZrBi is in good agreement with experimental values, as discussed earlier. The lowest and highest values of $\kappa$ are obtained for CoZrBi and FeTaBi, respectively. Interestingly, the calculated $\kappa$ values of Co-group are lower than those of Fe-group. Recently, $\kappa$ of CoZrBi is reported to be much lower than the existing hH alloys such as CoTiSb, CoZrSb, and FeNbSb \cite{Zhu18}. Our calculated $\kappa$ of CoTiBi and CoZrBi is lower and the $\kappa$ of CoHfBi almost parallel, in comparison to reported CoTiSb \cite{Wu07} and FeNbSb \cite{Tavassoli17}. 

Unfortunately, the $\kappa$ of FeNbBi and FeTaBi is higher than that of reported CoTiSb and FeNbSb. However, the $\kappa$ of FeVBi is close to the existing hH alloy CoTiSb. Importantly, the $\kappa$ of CoZrBi is almost twofold lower and that of CoTiBi is $\sim$1.3 times lower than the reported CoTiSb at room temperature. It is noteworthy that the $\kappa$ of reported CoZrBi shows a significant reduction at \textit{p}-type doping. Likewise, the other systems are supposed to have lower $\kappa$ on doping and thereby, the higher \textit{ZT} values. Incorporating the $\kappa$ of parent systems in \textit{ZT} formula (=$S^2\sigma T/\kappa$), next, we discuss the actual figures of transport coefficients and their analysis.

Table~\ref{zT} presents the optimal doping levels at which maximum PF is obtained. The corresponding values of PF, \textit{S}, and \textit{ZT} are also listed in Table~\ref{zT}. For comparison, the reported values of CoTiSb \cite{Wu07} and FeNbSb \cite{Fu15} are also included. The optimal doping levels range 0.20--0.25 hole doping per unit cell. The proposed doping levels are quite pragmatic and could be achieved experimentally. For instance, 0.20 hole doping can be achieved either by replacing 20\% Bi by Sn or 20\% Ta by Ti, etc. Wu \textit{et al.} \cite{Wu07} achieved high doping of 42\% in CoTiSb, i.e., Co$_{0.572}$Fe$_{0.428}$TiSb, whereas Zhu \textit{et al.} \cite{Zhu18} reported a doping of 35\% in CoZrBi. Hence, we believe that our proposed optimal doping levels are achievable. 

\begin{table*}
\caption{Calculated optimal doping levels and the corresponding Seebeck coefficient, electrical conductivity, power factor, and figure of merit of \textit{p}-type \textit{XYZ} half-Heusler alloys (\textit{X} = Co, Fe, \textit {Y} = Ti, Zr, Hf, V, Nb, Ta, and \textit{Z} = Bi) at 1100 K, assuming $\tau$ = 2 x 10$^{-15}$ s. The reported values of CoTiSb at 850~K and FeNbSb at~1100 K are taken as reference. \lq{+\rq} indicates the \textit{p}-type doping.}
\setlength{\arrayrulewidth}{0.5pt}
\begin{tabular*}{\textwidth}{c @{\extracolsep{\fill}} cccccc}
\hline
\hline
System		        &\textit{n} 	& S	    		  &$\sigma$		  &S$^2\sigma$		      &$ZT$  \\ 
                        & (e/uc)	&($\mu$V K$^{-1}$)	&(x 10$^{3}$ S cm$^{-1}$) &$\mu$W cm$^{-1}$ K$^{-2}$  &      \\ \hline		
CoTiSb \cite{Wu07}      & +0.15   	& 153       	          & -	          	  & 23.00            	      & 0.45 \\ \hline     
CoTiBi                  & +0.25        	& 170                     & 1.34                  & 39.06                     & 1.12 \\
CoZrBi                  & +0.24        	& 182                     & 1.24                  & 43.38                     & 1.53 \\ 
CoHfBi                  & +0.22        	& 162                     & 1.48                  & 39.26                     & 0.98 \\ \hline
FeNbSb \cite{Fu15}      & +0.20        	& 203                  	  & -	                  & 45.80                     & 1.1 \\ \hline
FeVBi                   & +0.23        	& 165                     & 1.38                  & 38.36                     & 0.81 \\ 
FeNbBi                  & +0.21        	& 159                     & 1.54                  & 39.20                     & 0.71 \\
FeTaBi                  & +0.20        	& 159                	  & 1.52                  & 38.67                     & 0.64 \\
\hline
\hline
\end{tabular*}
\label{zT}
\end{table*}

The Seebeck coefficient values range 159--182 $\mu$V K$^{-1}$ whereas the electrical conductivity values are in between 1.24--1.54$\times$10$^3$
S cm$^{-1}$. The calculated \textit{S} values of all the systems are higher than that of reported \textit{p}-type CoTiSb, however, the values are lower than the reported \textit{p}-FeNbSb. Nevertheless, the high Seebeck coefficient value of CoZrBi (182 $\mu$V K$^{-1}$) can be attributed to its high band degeneracy near the valence band maximum. Importantly, the PF of all the proposed systems is higher than that of reported CoTiSb. The highest PF value reported for \textit{p}-type CoTiSb is 23 $\mu$W cm$^{-1}$ K$^{-2}$ \cite{Wu07}. We find that our calculated values of PF of all the systems are $\sim$40\% higher than that of \textit{p}-type CoTiSb, which is a significant number. Remarkably, the PF of \textit{p}-type CoZrBi is $\sim$47\% higher than that of reported \textit{p}-type CoTiSb. The calculated PF values are not much behind from the reported \textit{p}-type FeNbSb. This suggests the thermoelectric potential of the proposed systems as \textit{p}-type FeNbSb is reported to have the highest \textit{ZT} among the hH alloys \cite{Yu17}. 

Focussing on the figure of merit, the obtained \textit{ZT} values at 1100~K range 0.71--1.53 which are again higher than that of reported \textit{p}-type CoTiSb. The maximum \textit{ZT} value obtained is 1.53 at 1100 K for \textit{p}-type CoZrBi, followed by 1.12 and 0.98 for \textit{p}-type CoTiBi and CoHfBi, respectively. The \textit{ZT} of \textit{p}-type CoTiBi and CoZrBi are higher than the reported \textit{p}-type FeNbSb. The \textit{ZT} values of Fe-group, as expected, are lower than that of Co-group. Nevertheless, the values are still a significant number. We stress that the proposed values are slightly underestimated on account of $\kappa$ of undoped systems. We believe that CoTiBi and CoHfBi are likely to be as promising as CoZrBi whereas the figure of merit of FeVBi and FeNbBi may reach unity, provided $\kappa$ of doped systems is considered.

We have successfully demonstrated that Bi-based hH alloys could be potential thermoelectric materials. As Bi-based hH alloys have not been much explored until recently, we hope that our findings will motivate further experimental studies. We are aware that theoretical predictions of new materials are relatively easier while experimental realization could be challenging as well as expensive. Therefore, both thermodynamic and dynamic stability of the proposed systems becomes crucial. Our calculations indicate the dynamic stability of all the proposed systems. According to the Open Quantum Materials Database (OQMD) \cite{Saal13, Kirklin15}, CoTiBi and CoHfBi are thermodynamically stable, however, the thermodynamic stability of FeVBi, FeNbBi, and FeTaBi is questionable. It should be noted that our proposed values are for \textit{p}-type doped systems instead of the parent composition. Although thermodynamic stability of parent composition suggest instability, the possibility of fabricating the doped systems cannot be completely ruled out. CoVSn is predicted to be thermodynamically unstable \cite{Zakutayev13, Saal13, Kirklin15}, however, Lue \textit{et al.} \cite{Lue01} experimentally shown that the material can be produced, at least with partial atomic ordering. Interestingly, they have shown the properties of CoVSn are closely related to the hH alloy. Therefore, we are of the opinion that optimizing the experimental conditions may be helpful in achieving the doped composition of Fe-group, if not the parent composition. 

\subsection{Summary}
Bismuth-based half-Heusler alloys CoTiBi, CoZrBi, CoHfBi, FeVBi, FeNbBi, and FeTaBi are theoretically investigated. The electronic features suggest the \textit{p}-type doping at Bi-site will be more favorable for obtaining high power factor values. At 700~K, FeTaBi shows similar power factor values on either \textit{p}-type or \textit{n}-type doping. Interestingly, at 1100~K, all the systems were found to have better power factor values than the reported \textit{p}-type CoTiSb and almost competitive to that of \textit{p}-type FeNbSb. The room temperature thermal conductivity of CoTiBi, CoZrBi, and CoHfBi is $\sim$13, 10, and 15 W m$^{-1}$ K$^{-1}$, respectively, in comparison to thermal conductivity of $\sim$17 W m$^{-1}$ K$^{-1}$ of CoTiSb. The maximum figure of merit obtained is 1.53 at 1100~K for \textit{p}-type CoZrBi, followed by 1.12 and 0.98 for \textit{p}-type CoTiBi and CoHfBi, respectively. We hope that our findings on bismuth-based half-Heusler alloys, suggesting a good thermoelectric potential, would serve as a base for experimental study. 

\subparagraph{Acknowledgments}
S.S. and M.Z. are thankful to MHRD and CSIR, respectively, for the support of a senior research fellowship. Computations were performed
at IFW Dresden, Germany. We thank Ulrike Nitzsche for technical assistance.

\end{document}